\documentclass[%
prd
 ,twocolumn%
 ,secnumarabic%
,amssymb, amsmath,nobibnotes,showpacs]{revtex4}
\usepackage{bm}%
\expandafter\ifx\csname package@font\endcsname\relax\else
 \expandafter\expandafter
 \expandafter\usepackage
 \expandafter\expandafter
 \expandafter{\csname package@font\endcsname}%
\fi

\begin{document}

\title{Matter in Loop Quantum Gravity without time gauge: a non-minimally coupled scalar field}%

\author{Francesco Cianfrani$^{1}$, Giovanni Montani$^{123}$}%
\email{montani@icra.it, francesco.cianfrani@icra.it}
\affiliation{$^{1}$ICRA-International Center for Relativistic Astrophysics, Dipartimento di Fisica (G9),\\
Universit\`a  di Roma, ``Sapienza'', Piazzale Aldo Moro 5, 00185 Roma, Italy.\\ 
$^{2}$ENEA C.R. Frascati (Dipartimento F.P.N.), Via Enrico Fermi 45, 00044 Frascati, Roma, Italy.\\
$^{3}$ICRANet C. C. Pescara, Piazzale della Repubblica, 10, 65100 Pescara, Italy.}
\date{July 2009}%

\begin{abstract}

We analyze the phase space of gravity non-minimally coupled to a scalar field in a generic local Lorentz frame. We reduce the set of constraints to a first-class one by fixing a specific hypersurfaces in the phase space. The main issue of our analysis is to extend the features of the vacuum case to the presence of scalar matter by recovering the emergence of an SU(2) gauge structure and the non-dynamical role of boost variables. Within this scheme, the super-momentum and the super-Hamiltonian are those ones associated with a scalar field minimally coupled to the metric in the Einstein frame. Hence, the kinematical Hilbert space is defined as in canonical Loop Quantum Gravity with a scalar field, but the differences in the area spectrum are outlined to be the same as in the time-gauge approach. 
\end{abstract}


\maketitle

\section{Introduction} 
Loop Quantum Gravity (LQG) is the most promising approach for a non-perturbative description of the quantum gravitational field \cite{revloop}. Indeed, the implementation of the dynamics and of a proper semi-classical limit are still open issues.  However, the main progress with respect to the Wheeler-DeWitt formulation \cite{DW} is the definition of a proper kinematical Hilbert space \cite{LOST06}, where all constraints but the super-Hamiltonian one can be solved and the spectra of geometrical operators turn out to be discrete \cite{discr}, as expected for a quantum geometry. The LQG approach is based on a reformulation of gravity in terms of real SU(2) connections (Ashtekar-Barbero-Immirzi connections) and on the quantization of the corresponding holonomy-flux algebra. 

The origin of this $SU(2)$ gauge symmetry was traced back to the assignment of a fixed local Lorentz frame, realized via the time-gauge condition. The development of a covariant LQG formulation was performed in \cite{ale} (see also \cite{liv}). In these works the time gauge condition was relaxed and a Lorentz connection $A^X_i$ could be defined. The quantization was performed in terms of holonomies of the SO(1,3) group and by a suitable projection (which depends on boost parameters, $\chi_a$) it was shown that the kinematical Hilbert space of LQG, whose basis vectors are invariant SU(2) spin-networks, appeared. 

In \cite{prl}, a classical formulation free of any $\chi$-dependence was achieved. In fact, it has been demonstrated that the constraints associated with the local Lorentz symmetry in the vacuum case reduce to those ones of a true SU(2) gauge theory, independently of fixing specific boost parameters. Other conditions just ensure the non-dynamical role of $\chi_a$. This new set of constraints can be safely implemented on a quantum level and the corresponding Hilbert space is the same as in LQG for any value of $\chi_a$.
 
In this work, we extend this result to the case in which a non-minimally coupled scalar field is present. This case was already considered in \cite{MM}, where no gauge fixing of the local Lorentz frame was performed, but complex connections were considered. Then, in \cite{AC03} the quantization was performed with real connections but within the time gauge.  

The interest for a model with gravity and a scalar field is enforced by the fact that such a dynamical system describes $f(R)$ theories (see \cite{fr} for a review). These modified gravity models can account for un-explained features of the Universe in a cosmological setting. 

Here we perform the analysis of the Hamiltonian constraints emerging in presence of gravity and a non-minimally coupled scalar field without the simplifications allowed by the choice of the time gauge. The non-minimal coupling introduces a non-trivial interaction between the geometry and the scalar field, thus making the development of the Hamiltonian formulation without the time gauge an interesting issue. In particular we demonstrate that, as soon as 3-bein vectors and the scalar field in the Einstein frame are identified as proper phase-space variables, the analysis of constraints can be performed as in vacuum \cite{prl}, thus variables describing the local Lorentz frame are non-dynamical. The main issue of our analysis is to outline that, like in \cite{AC03}, the super-Hamiltonian and the super-momentum associated with a minimally-coupled scalar field till appear, but the kinematical quantization of the model takes now place without regarding the choice of the local Lorentz frame. In fact, the kinematical Hilbert space associated with such a system can be properly defined as when the time gauge holds \cite{Thmatt}, because the Gauss constraints retain the proper form they have in the vacuum and time-gauge case.

The complete equivalence of our generalized case with the analysis of \cite{AC03} enforces the idea that the non-minimally coupling nature of the scalar field be, in the loop formulation, just the counterpart of a minimally-coupled scalar field able to affect the quantum geometry scaling. Thus we get the surprising result that, in this scheme, it is the matter to influence the microscopic structure of the space time, instead of the local Lorentz frame in which it is looked.

As for $f(R)$ theories, one can reduce such dynamical system to Einstein-Hilbert gravity with a non-minimally coupled scalar field. Hence, the independence of the quantum formulation from a particular gauge fixing of the local Lorentz frame is established. Furthermore, it is outlined that the scalar field enters also in this case into the definition of the scale at the which the discreteness of the spatial geometry arises.

The manuscript is organized as follows: section \ref{0} is devoted to discuss the differences between Covariant Loop Quantum Gravity and Loop Quantum Gravity without the time gauge, while in section \ref{1} the latter is extended to the case a non-minimally coupled scalar field is present. In particular, the Hamiltonian structure is presented and the set of constraints is recognized as being second-class. Then, the whole set is reduced to a first-class one by restricting to a specific hypersurfaces of the phase space. The analysis of constraints on such an hypersurfaces is performed and the emergence of SU(2) Gauss constraints is outlined. Hence, the kinematical Hilbert space is defined and the form of area eigen-values is discussed. In section \ref{2} the relevant features of the loop quantization for f(R) theories are illustrated. Brief concluding remarks follow in section \ref{5}.

\section{Covariant Loop Quantum Gravity vs Loop Quantum Gravity without the time gauge}\label{0}

The first approach towards a non-perturbative quantization of gravity without any local Lorentz gauge fixing is due to Alexandrov \cite{ale}. In his works, second class constraints, which appeared when the time gauge is relaxed \cite{BS},  were solved by replacing Poisson with Dirac brackets and a Lorentz connection $A^X_i$ was defined. 

The quantization was performed by defining holonomies and fluxes associated with the SO(1,3) group, while the scalar product could be developed from the corresponding Haar measure. Nevertheless, in view of implementing consistently the second-class constraints on a quantum level, spin networks had to be replaced by projected spin networks. The latter were developed by inserting suitable projectors at each vertex and they were functions of connections associated to a modified  $SU(2)$ group, $SU(2)_\chi$. Finally, such a basis was shown to coincide with that of $SU(2)$ spin networks. Therefore, Covariant LQG reproduced the Hilbert space structure proper of standard LQG. 
 
The procedure adopted in \cite{prl} was different on a formal level and it elucidated the link between a covariant theory and an SU(2) one. In particular, second-class constraints were solved explicitly at the classical level and the boost constraints were reduced to the rotational ones. No gauge fixing was performed since $\chi_a$ were promoted to dynamical variables. Within this scheme 
SU(2) connections, $\widetilde{A}^a_i$, could be inferred and conjugate momenta to $\chi_a$ were constrained to vanish. Hence, the kinematical Hilbert space could be developed by disregarding any $\chi$-dependence from physical states and $SU(2)$ spin-networks realized a basis. Therefore, the kinematical Hilbert space was the same as when the time gauge holds and no local gauge condition was fixed.  

Indeed, $\widetilde{A}^a_i$ can be recognized to be the non-vanishing components of Alexandrov covariant ones $A^X_i$ after a proper $\chi$-dependent Lorentz transformation (as outlined in a private communication with Alexandrov). The merit of the paper \cite{prl} was to point out that SU(2) connections are privileged functions in the classical formulation, since they contain the full dynamical information.  

\section{Non-minimally coupled scalar field}\label{1}

\subsection{Hamiltonian structure}

Let us consider the Holst action \cite{Ho96} with a non-minimally coupled scalar field $\phi$, {\it i.e.} (in units $c=8\pi G=1$) 
\begin{eqnarray}  
S=\int \sqrt{-g}\bigg[F(\phi)e^\mu_A e^\nu_B R_{\mu\nu}^{CD}(\omega^{FG}_\mu){}^\gamma\!p^{AB}_{\phantom1\phantom2CD}+\nonumber\\+\frac{1}{2}g^{\mu\nu}K(\phi)\partial_\mu\phi\partial_\nu\phi-V(\phi)\bigg]d^4x,\label{actnm}
\end{eqnarray}

$g_{\mu\nu}$ being the metric tensor, whose 4-bein vectors and spinor connections are $e^A_\mu$ and $\omega^{AB}_\mu$, respectively, while $R^{AB}_{\mu\nu}$ denotes the expression
\begin{equation}
R^{AB}_{\mu\nu}=2\partial_{[\mu}\omega^{AB}_{\nu]}-2\omega^A_{\phantom1C[\mu}\omega^{CB}_{\nu]}.
\end{equation}

We also consider a non-standard kinetic term, by taking an arbitrary function $K(\phi)$.

As for ${}^\gamma\!p^{AB}_{\phantom1\phantom2CD}$, it contains the Immirzi parameter $\gamma$ as follows 
\begin{equation}
{}^\gamma\!p^{AB}_{\phantom1\phantom2CD}=\delta^{AB}_{\phantom1\phantom2CD}-\frac{1}{2\gamma}\epsilon^{AB}_{\phantom1\phantom2CD}. \label{EH}
\end{equation}

The function $F(\phi)$ introduces a non-trivial interaction between the geometry and the scalar field. In particular, if we fix $F(\phi)=1+\xi\phi^2$, then the parameter $\xi$ gives the amount of the non-minimal coupling between the geometry and the scalar field. 
Upon solving equations of motion for $\omega^{AB}_i$, the action (\ref{actnm}) can be shown to be equivalent to the second-order one for gravity \cite{ACS03}.

We take $\{\omega^{AB}_i,\phi\}$ as configuration variables and their conjugate momenta turn out to be
\begin{equation} 
{}^\gamma\!\pi_{AB}^i={}^\gamma\!p^{CD}_{\phantom1\phantom2AB}\pi_{CD}^i,\qquad {}^\phi\!\pi=K(\phi)(\sqrt{-g}g^{tt}\partial_t\phi+\sqrt{-g}g^{ti}\partial_i\phi). 
\end{equation}

In the equation above we introduced the expressions $\pi^i_{AB}=2F(\phi)\sqrt{-g}e^t_{[A}e^i_{B]}$. As soon as the ADM splitting of the space-time manifold is provided, those quantities allow us to establish a direct link between phase-space variables and the 3-metric $h_{ij}$, since the following relations hold
\begin{equation}
hh^{ij}=\frac{1}{2F^2}\eta^{AC}\eta^{BD}\pi^i_{AB}\pi^j_{CD}.
\end{equation}
  
The Hamiltonian density is given by the following linear combination of constraints
\begin{eqnarray}   \mathcal{H}=\int\bigg[\frac{\widetilde{N}}{\sqrt{h}}H+\widetilde{N}^iH_i-\omega^{AB}_t{}^\gamma\!p^{CD}_{\phantom1\phantom2AB}G_{CD}+\nonumber\\+
\lambda_{ij}C^{ij}+\eta_{ij}D^{ij}+\lambda^{AB}\pi_{AB}^t\bigg]d^3x,\end{eqnarray}

where $\widetilde{N}$ and $\widetilde{N}^i$ denote the lapse function and the shift vector, respectively, while ${}^\gamma\!p^{CD}_{\phantom1\phantom2AB}\omega^{AB}_t$, $\lambda_{ij}$, $\eta_{ij}$ and $\lambda^{AB}$ are other Lagrangian multipliers. In particular we have
\begin{itemize}
{\item the super-Hamiltonian constraint, which reads
\begin{eqnarray}   
H=\frac{1}{F}\pi^i_{CF}\pi^{jF}_{\phantom1D}{}^\gamma\!p^{CD}_{\phantom1\phantom2AB}R^{AB}_{ij}+\frac{1}{2K}{}^\phi\!\pi^2+\nonumber\\+\frac{K}{4F^2}\pi^i_{AB}\pi^{jAB}\partial_i\phi\partial_j\phi+hV(\phi)=0,\label{sh} 
\end{eqnarray}

and it enforces the invariance under parametrization of the time variable;}
{\item the super-momentum constraints, {\it i.e.}
\begin{equation}
H_i={}^\gamma\!p_{AB}^{\phantom1\phantom2CD}\pi^j_{CD}R^{AB}_{ij}+\pi\partial_i\phi=0,\label{sm}
\end{equation}

which generate 3-diffeomorphisms;}
{\item the Gauss constraints associated with the local Lorentz transformations, which can be written as
\begin{equation}
G_{AB}=D_i\pi^i_{AB}=\partial_i\pi^i_{AB}-2\omega_{[A\phantom2i}^{\phantom1C}\pi^i_{|C|B]}=0;\label{LG}
\end{equation}
}
{\item the following additional conditions
\begin{eqnarray}
C^{ij}=\epsilon^{ABCD}\pi_{AB}^{(i}\pi_{CD}^{j)}=0,\label{C} \\ D^{ij}=\epsilon^{ABCD}\pi^k_{AF}\pi^{(iF}_{\phantom1\phantom2B}D_k\pi^{j)}_{CD}=0,
\label{D}
\end{eqnarray}
with the latter coming out as secondary constraints from the former.
}
\end{itemize}

Since $\{C^{ij},D^{kl}\}$ and $\{D^{ij},D^{kl}\}$ do not vanish on shell, the set of constraints (\ref{sh}), (\ref{sm}), (\ref{LG}), (\ref{C}) and (\ref{D}) is second-class.

\subsection{Solution of second-class constraints}

It is worth noting that the constraints (\ref{C}) and (\ref{D}) coincide with the ones in vacuum, such that they can be solved as in \cite{prl}.

In particular, we parametrize the solutions in the following way 
\begin{eqnarray}
\pi^i_{ab}=2\chi_{[a}\pi^i_{b]},\\ \omega^{\phantom1b}_{a\phantom1i}={}^\pi\!\omega^{\phantom1c}_{a\phantom1i}T^{-1b}_c+\chi_a\omega^{0b}_{\phantom{12}i}+\chi^b
(\omega_{a\phantom1i}^{\phantom10}-\partial_i\chi_a).\label{scon}
\end{eqnarray}

where $\pi^i_a=\pi^i_{0a}$ and ${}^\pi\!\omega^{\phantom1b}_{a\phantom1i}=\frac{1}{\pi^{1/2}}\pi^b_l{}^3\!\nabla_i(\pi^{1/2}\pi^l_a)$. Here ${}^3\!\nabla_i$ denote the covariant derivatives associated with the fictitious 3-metric (see below) ${}^\phi\!h_{ij}=-\frac{1}{\pi}T^{-1}_{ab}\pi^a_i\pi^b_j$, with $\pi$ the determinant of $\pi^a_i$ and $T^{-1}_{ab}=\eta_{ab}+\chi_a\chi_b$.

As for $\chi_a$ they are the variables labeling the local Lorentz frame, which is given by the following set of 4-bein vectors
\begin{eqnarray}
e^0=Ndt+\chi_a E^a_idx^i,\qquad e^a=E^a_iN^idt+E^a_idx^i.
\end{eqnarray} 

These 4-bein components are related with the phase space coordinates, in fact for the lapse function and the shift vector we have
\begin{equation}
\widetilde{N}=\frac{N-N^i\chi_aE^a_i}{\sqrt{1+\chi^2}},\qquad \widetilde{N}^i=N^i+\frac{N-N^i\chi_bE^b_i}{1+\chi^2}\chi^aE_a^i,
\end{equation}

where $\chi^a=\eta^{ab}\chi_b$ and $\chi^2=\chi^a\chi_a$, while the 3-metric $h_{ij}$ is given by the expression
\begin{equation}
h_{ij}=-T^{-1}_{ab}E^a_iE^{b}_j,\qquad E^a_i=\frac{1}{\sqrt{h}(1+\chi^2)F}\pi_i^a. 
\end{equation}

The fictitious 3-metric ${}^\phi\!h_{ij}$ can be obtained from the real 3-metric $h_{ij}$ by
\begin{equation}
{}^\phi\!h_{ij}=F(\phi)h_{ij}, 
\end{equation}
 
and its inverse densitized 3-bein vectors $\widetilde{\pi}^i_a$ read as follows
\begin{equation}
\widetilde{\pi}^i_a=S^b_a\pi^i_b,\qquad S_a^b=\sqrt{1+\chi^2}\delta^a_b+\frac{1-\sqrt{1+\chi^2}}{\chi^2}\chi_a\chi_b.
\end{equation} 

It is worth noting that ${}^\phi\!h_{ij}$ is the 3-metric in the Einstein frame, but it has to be regarded as fictitious in comparison to the real geometrical one $h_{ij}$ in the Jordan frame.

\subsection{Analysis of Hamiltonian constraints}\label{3}

The analysis of constraints (\ref{sh}), (\ref{sm}) and (\ref{LG}) with conditions (\ref{scon}) can be performed as in vacuum \cite{prl}, once replacing $h_{ij}$ with ${}^\phi\!h_{ij}$.

Therefore, we take $\{\widetilde{\pi}^i_a,\chi_a\}$ as configuration variables, while corresponding conjugate momenta are given by $\widetilde{A}^a_i$ and $\widetilde{\pi}^a$. $\widetilde{A}^a_i$ are generalized Ashtekar-Barbero-Immirzi connections, both with respect to the removal of the time gauge and the presence of the scalar field, and their expression is given by 
\begin{eqnarray}
\widetilde{A}_i^a=S^{-1a}_b\bigg((1+\chi^2)T^{bc}(\omega_{0ci}+{}^\pi\!D_i\chi_c)-\nonumber\\-\frac{1}{2\gamma}\epsilon^b_{\phantom1cd}{}^\pi\!\omega^{cf}_{\phantom1\phantom2i}T^{-1d}_{\phantom1f}+\frac{2+\chi^2-2\sqrt{1+\chi^2}}{2\gamma\chi^2}\epsilon^{abc}\partial_i\chi_b\chi_c\bigg).\quad\label{ABI}
\end{eqnarray}

Hence, according to the analysis developed in \cite{prl}, the following constraints are inferred from $G_{AB}=0$
\begin{equation}
G_a=\partial_i\widetilde{\pi}^i_a+\gamma\epsilon_{ab}^{\phantom{12}c}\widetilde{A}^b_i\widetilde{\pi}_c^i=0,\qquad \widetilde{\pi}^a=0.\label{con1}
\end{equation}

The former are Gauss constraints of the SU(2) group, which, by virtue of the general frame we adopted, emphasize the crucial role of such a gauge group in the Hamiltonian formulation for gravity as in the vacuum case. We stress how such a constraint is stated with explicit dependence on the $\phi$ variables (indeed $\phi$ enters the definition of momenta in terms of the geometrical 3-beins $E^a_i$). It is just this feature which will allow in the next section to deal with a standard approach for the kinematics of LQG.

The latter give the vanishing of momenta conjugate to $\chi_a$ and they outline that such variables do not play any dynamical role. 

It is worth noting that we were able to reduce the full Lorentz-Gauss constraints to the two set of independent conditions (\ref{con1}). This is a key point towards quantization, since when dealing with a non-compact gauge group (as the Lorentz one) some divergences would have arisen performing the integration on the group manifold via the Haar measure. Instead here conditions (\ref{con1}) can be safely implemented on a quantum level by standard techniques for the compact SU(2) group. 

Other constraints are the super-momentum and the super-Hamiltonian ones. In particular by a redefinition of the Lagrangian multipliers in front of $H$, the super-Hamiltonian can be multiplied times $F(\phi)$, such that by performing the canonical transformation 
\begin{equation}
\phi\rightarrow\varphi=\int^\varphi F^{-1/2}(\phi)d\phi,\qquad{}^\phi\!\pi\rightarrow{}^\varphi\!\pi=F^{1/2}(\phi){}^\phi\!\pi, 
\label{tr}\end{equation}

one finds the Hamiltonian constraints associated with a minimally coupled scalar field $\varphi$, {\it i.e.} 
\begin{eqnarray}
H'=\pi^i_{CF}\pi^{jF}_{\phantom1D}{}^\gamma\!p^{CD}_{\phantom1\phantom2AB}R^{AB}_{ij}+\frac{1}{2K}{}^\varphi\!\pi^2+\nonumber\\+\frac{K}{4}{}^\phi\!h{}^\phi\!h^{ij}\partial_i\varphi\partial_j\varphi+\frac{{}^\phi\!h}{F(\phi(\varphi))^2}V\left(\phi(\varphi)\right)=0,\quad\\
H_i={}^\gamma\!p_{AB}^{\phantom1\phantom2CD}\pi^j_{CD}R^{AB}_{ij}+{}^\varphi\!\pi\partial_i\varphi=0.\quad
\end{eqnarray}

Therefore, it is possible to describe the dynamics of gravity non-minimally coupled to a scalar field by that of a ``fake'' geometry, whose 3-metric is ${}^\phi\!h_{ij}$, and a minimally coupled scalar field.

The transformation (\ref{tr}) is the analogous in the phase space of the transition between the Jordan and Einstein frame in the Lagrangian. However, the SU(2) gauge constraints are inferred only for momenta $\widetilde\pi^i_a$, which are densitized inverse 3-bein vectors of the fictitious metric ${}^\phi\!h_{ij}$. Hence the LQG quantization procedure works for geometric variables of the Einstein frame. 

\subsection{Loop quantization of the model}\label{4}

The canonical quantization can be performed along the lines of the standard LQG in presence of a minimally-coupled scalar field. 
In particular, the holonomy of the SU(2) group can be defined for connections $\widetilde{A}^a_i$ and the corresponding holonomy-flux algebra can be quantized. The kinematical Hilbert space is given by the direct product of 
\begin{itemize}
{\item the one proper of LQG, whose basis are invariant spin-networks and with the Ashtekar-Lewandoski measure \cite{ALMMT95};} 
{\item the one corresponding to the scalar field on a background independent theory, taking point-like holonomies $U_\varphi(x)=e^{i\varphi}$ as basic variables and the associated Ashtekar-Lewandoski measure (see \cite{Thmatt} for the explicit construction).} 
\end{itemize}
 
Within this scheme, the vanishing of $\widetilde{\pi}^a$ can be implemented taking wave-functionals not depending on $\chi_a$. This point emphasizes that those variables do not play any dynamical role on a quantum level, too. 

If we investigate the area operator of a surface $S$, we find the standard discretized spectrum for the area associated to ${}^\phi\!h_{ij}$, while the true area operator contains a factor $F$ as follows
\begin{equation}
A(S)h_e(\widetilde{A})=\gamma \frac{l_p^2}{F(\phi)} Ch_e(\widetilde{A}),\label{area}
\end{equation}

$h_e(\widetilde{A})$ being the parallel transport of $\widetilde{A}^a_i$ along an edge $e$ passing through $S$, while $C=\delta^{ab}\tau_a\tau_b$ is the Casimir operator of the $SU(2)$ group, whose generators are $\tau_a$. Here, $l_P$ is the Planck length, which arises when $\hbar$, $G$ and $c$ are reinserted. 

Hence, like in \cite{AC03}, the area eigen-values depend on the field $\varphi=\varphi(\phi)$ and this result emphasizes how the presence of a non-minimally coupled scalar field determines the scale at which the geometry outlines its discrete nature. 

The same result was obtained within the time gauge in \cite{AC03} and our analysis demonstrates that it does not depend on the adopted gauge fixing of the local Lorentz frame. This issue has the significant feature that the boost functions $\chi_a$ does not enter the spectra of geometrical operators, unlike the scalar field does. Therefore, the non-trivial matter-geometry coupling, fixed by the analysis in \cite{AC03}, cannot be removed by the choice of a specific local Lorentz frame and thus it is an intrinsic physical feature.

\section{Modified gravity}\label{2}

It is interesting to trace an analogy between the way a non-minimally coupled scalar field enters into the definition of the discrete spatial structure with a modified gravity theory. In particular, $f(R)$ theories are based on replacing the Einstein-Hilbert action with one whose Lagrangian is a generic function $f$ of the scalar curvature, {\it i.e.}
\begin{equation}
S_{MOD}=\int d^4x\sqrt{-g}f(R^{ab}_{\mu\nu}e^\mu_ae^\nu_b).\label{MOD}
\end{equation}

A standard tool in such a model \cite{fr} consists in introducing the scalar mode $\phi$, such that the action (\ref{MOD}) can be rewritten as
\begin{equation}
S_{MOD}=\int d^4x\sqrt{-g}[\phi R^{ab}_{\mu\nu}e^\mu_ae^\nu_b+V(\phi)],
\end{equation}

where the potential term reads
\begin{equation}
V=\phi (f')^{-1}(\phi)-f((f')^{-1}(\phi)).
\end{equation}

An equivalent description can be given at the first order taking the following action  
\begin{eqnarray}
S_{MOD}=\int d^4x\sqrt{-g}\bigg(\phi R^{CD}_{\mu\nu}e_A^\mu e_B^\nu{}^\gamma\!p^{AB}_{\phantom{12}CD}+\nonumber\\+\frac{3}{2\phi}g^{\mu\nu}\partial_\mu\phi\partial_\nu\phi-V(\phi)\bigg).
\end{eqnarray}

The introduction of the additional scalar field $\phi$ allows us to described the $f(R)$ model as a Brans-Dicke theory \cite{BD}. This means that as soon as the scalar-tensor representation is addressed, $\phi$ is treated as an external matter field. 

Therefore, the analysis of the Hamiltonian constraints and the loop quantization can be carried on as discussed previously, after setting $F(\phi)=\phi$ and $K(\phi)=3/\phi$ into the action (\ref{actnm}). 

Hence, the area operator acts on holonomies as follows
\begin{equation}
A(S)h_e(\widetilde{A})=\gamma\frac{l_p^2}{\phi} Ch_e(\widetilde{A}),
\end{equation} 

where the notation is the same as in equation (\ref{area}).

The relation above points out that the scalar field enters into the definition of the discrete structure of the spatial geometry. This result is impressive and it reminds us that $\phi$ is not a true external scalar field, but it arises as a consequence of the new dynamical features proper of the geometry in the $f(R)$ models.  

However, we remark that the area operator retains the same form both in the original $f(R)$ theory as in the first-order Brans-Dicke model we quantized here. Thus to split the original geometrodynamics in terms of redefined gravitational degrees of freedom and a matter source can be regarded as an algorithm to infer how the fundamental space-time has a discrete structure.

Therefore the conclusions we got here allow to claim that a general $f(R)$ model can be addressed in a Lorentz-invariant formulation when its scalar-tensor representation is implemented. In this picture, we deal with a Lorentz-invariant LQG theory whose geometrical observables are influenced by the presence of the original scalar degrees of freedom, contained in the form of the Lagrangian. 

\section{Conclusions}\label{5}
The Hamiltonian formulation of gravity with a non-minimally coupled scalar field has been performed without fixing the local Lorentz frame. As soon as basic phase space variables were recognized as the ones associated with the (fictitious) 3-metric in the Einstein frame, the analysis of constraints was the same as in the vacuum case. Hence, SU(2) Gauss constraints were inferred, such that the holonomy-flux algebra could be quantized. Then, the vanishing of conjugate momenta to $\chi_a$ ensured that physical states did not depend on the local Lorentz frame.    

Therefore, the introduction of a scalar field does not modify the conclusions of \cite{prl}. It is possible to extend the LQG quantization procedure to a generic local Lorentz frame and the invariance of geometrical operator spectra from variables labeling the frame is obtained. 

This result is an encouraging starting point in view of demonstrating that the LQG formulation is not affected by the choice of a specific Lorentz frame, even when the coupling with an external matter field is concerned. In this respect, the non-minimally coupled scalar field we addressed above has the relevant feature to enter the Ashtekar-Barbero-Immirzi connections and, hence, the SU(2) constraints. All matter fields which does not single out such a feature would automatically preserve the analysis presented in \cite{prl} as far as the Lorentz frame is regarded.

Then, it was discussed the possibility to quantize a vacuum $f(R)$ theory in the Einstein frame, where the dynamics coincided with the Einstein-Hilbert one in presence of a non-minimally coupled scalar field $\phi$. The possibility to relax the time-gauge condition has been already established by the previous analysis, while it has been emphasized that the discrete structure arising after the loop quantization was affected by a $\phi$-dependent term.

\end{document}